\documentclass[page-classic]{epl2}

\title{Classical no-cloning theorem under Liouville dynamics by non-Csisz{\' a}r f-divergence}
\shorttitle{Title} 

\author{T. Yamano\inst{1}\thanks{E-mail: \email{yamano@amy.hi-ho.ne.jp}} 
\and O. Iguchi\inst{1}\thanks{E-mail: \email{osamu@phys.ocha.ac.jp}}}
\shortauthor{T. Yamano \etal}

\institute{                    
\inst{1} Department of Physics, Ochanomizu University, 2-1-1 Otsuka, Bunkyo-ku, 
Tokyo 112-8610, Japan\\
}

\abstract{
The Csisz{\' a}r f-divergence, which is a class of information distances, is known to offer a useful tool for 
analysing the classical counterpart of the cloning operations that are quantum mechanically impossible for 
the factorized and marginality classical probability distributions under Liouville dynamics. We show that 
a class of information distances that does not belong to this divergence class also allows for the formulation 
of a classical analogue of the quantum no-cloning theorem. We address a family of nonlinear Liouville-like 
equations, and generic distances, to obtain constraints on the corresponding functional forms, associated with 
the formulation of classical analogue of the no-cloning principle. 
}

\begin{document}

\maketitle

\section{Introduction}
The search for classical analogues of the processes that are quantum mechanically impossible allows 
one to identify those aspects that are purely quantum mechanical. Along this line, a classical analogue 
of the quantum entanglement has been discussed, namely, secret classical correlations \cite{Collin}.
Moreover, it has been reported that principles of quantum no-cloning \cite{Woot,Dieks} and no-deleting 
operations \cite{Pati} under unitary dynamics possess classical analogues \cite{CNCT}. The existence of 
parallelism between classical and quantum copying processes has been demonstrated via 
contradictions between the behaviors of relative entropies. That is, the relative entropy calculated 
from statistical ensemble distributions changes upon copying, which is in conflict with the time 
invariance of that measure under Liouville dynamics.\\

The relative entropy of the probabilities defined by Kullback and Leibler 
$D_{KL}(\mathcal{P}_1,\mathcal{P}_2)=\int dx\mathcal{P}_1\ln(\mathcal{P}_1/\mathcal{P}_2)$ is constant 
in time when the probabilities obey the common Liouville dynamics \cite{Mac,CNCT}. 
This implies that the Kullback-Leibler distance, i.e., the relative entropy, remains the same as 
that given by the initial probabilities with which the original system is prepared. The Csisz{\' a}r 
f-divergence \cite{Cs} measures the distance between two probability distributions $\mathcal{P}_1$ 
and $\mathcal{P}_2$ by the form $\mathcal{F}=\int dx\mathcal{P}_1f(\mathcal{P}_2/\mathcal{P}_1)$, 
where $f$ is a convex function satisfying $f(1)=0$. Then, the Kullback-Leibler distance is found 
to be a special case of the f-divergence ($f=-\ln[\mathcal{P}_2/\mathcal{P}_1]$, \cite{KL}).\\

As an extended proof of the classical no-cloning theorem along the line in \cite{CNCT}, 
Plastino and Daffertshofer \cite{Plastino} have showed that the same contradictory situation upon 
copying a source system arises when employing a wider class of distance measures, i.e., the Csisz{\' a}r 
f-divergence. Under the Liouville dynamics, the conservation of the f-divergence is incompatible 
with a change in the distance during the copying process, where the following three different reasonable 
premises for the input and output distributions are made:\\ 
(i) For the initial distributions, the state of a system is expressed by a product of distribution 
functions of three components as $\mathcal{P}_j=\mathcal{P}_j^{(o)}\mathcal{P}^{(a)}\mathcal{P}^{(r)}$. 
This reads as follows: the state of the original system $(o)$ is copied to the target system (replica) 
$(r)$ \cite{comm} with the help of an ancillary copying machine $(a)$. The index $j$ specifies the 
different initial conditions. The coordinates $\vec{x}^{(\cdot)}$ comprise the support of each 
distribution, i.e., $\mathcal{P^{(\cdot)}}(\vec{x}^{(\cdot)})$. This implies that the original system 
to be copied is prepared in such a way that its state is statistically independent of the machine 
and the replica.\\
(ii) The successful cloning in (i) yields the final distribution as 
$\mathcal{Q}_j=\mathcal{P}_j^{(o)}\mathcal{Q}^{(a)}\mathcal{P}_j^{(o)}$. During the copying process, 
the machine interacts with both the original system and its replica, and it functions as a reader and 
a writer. As a result, it changes its states from $\mathcal{P}^{(a)}$ to $\mathcal{Q}^{(a)}$.\\
(iii) The marginalization for the final distribution $\mathcal{Q}_j$ with respect to 
the coordinates of the machine is used: 
$\int d\vec{x}^{(a)}\mathcal{Q}_j=\mathcal{P}_j^{(o)}(\vec{x}^{(o)})\mathcal{P}_j^{(o)}(\vec{x}^{(r)})$.\\

With this background, we ask in this Letter: is the proof of the existence 
of the classical no-cloning theorem specific for the choice of the relative entropy? There may be other 
relative distance measures that also display a conflict under Liouville dynamics. To answer this question we 
investigate the case of non f-divergence-type distances with which, if valid, we can provide robustness 
to the delineation of the classical counterpart of the quantum process. We show that the combination of 
the f-divergence-type distance for measuring probabilities 
and the Liouville dynamics is a strong requirement for establishing the classical counterpart. 
We exemplify it by considering non f-divergence distances under a Liouville type equation 
incorporating a power law nonlinearity. In the next section, we first introduce a simple class of distance 
measures that do not belong to the f-divergence class. In section III, we show the discrepancy that leads 
to the equivalent conclusion given in \cite{CNCT,Plastino}. In section IV, we consider nonlinearities in 
the Liouville equation for two types of distance measures, before closing with a summary and concluding 
remarks.
\section{A distance of non-Csisz{\' a}r f-divergence type}
We consider the continuous Chernoff $\alpha$-distance \cite{Cher} between two distinct probability 
distributions under phase space dynamics $\vec{x}_t$, 
\begin{equation}
C_\alpha(\mathcal{P}_1(\vec{x}_t),\mathcal{P}_2(\vec{x}_t))\equiv 
-\log\left[ \int d\vec{x}_t\mathcal{P}^\alpha_1(\vec{x}_t)\mathcal{P}^{1-\alpha}_2(\vec{x}_t)\right]
\label{eqn:defC}.
\end{equation}
We exclude the situation $\mathcal{P}_1^\alpha\mathcal{P}_2^{1-\alpha}=0$, that is, the two 
distributions do not have a common support, where the distance is not well defined. 
From an inequality $a^\alpha b^{1-\alpha}\geq {\rm min}\{a,b\}$ for $(0\leq \alpha \leq 1)$, 
the positivity $C_\alpha (\mathcal{P}_1,\mathcal{P}_2)\geq 0$ is assured allowing for using it 
as a distance. Similarly, it is asymmetric with respect to $\mathcal{P}_1$ and $\mathcal{P}_2$ for 
general value of $\alpha$. The distance is zero if and only if $\mathcal{P}_1=\mathcal{P}_2$ holds 
for a fixed value of $\alpha$. When $\mathcal{P}_1\neq\mathcal{P}_2$, 
then $C_0(\mathcal{P}_1,\mathcal{P}_2)=C_1(\mathcal{P}_1,\mathcal{P}_2)=0$ holds. The Kullback-Leibler 
distance is generated from the differential coefficient, e.g., 
$D_{KL}(\mathcal{P}_1,\mathcal{P}_2)=-[dC_\alpha(\mathcal{P}_1,\mathcal{P}_2)/d\alpha]_{\alpha=1}$ and 
$D_{KL}(\mathcal{P}_2,\mathcal{P}_1)=[dC_\alpha(\mathcal{P}_1,\mathcal{P}_2)/d\alpha]_{\alpha=0}$. 
A particular choice $\alpha=1/2$ provides the Bhattacharyya distance \cite{Bha},   
$C_{1/2}(\mathcal{P}_1,\mathcal{P}_2)=-\log\left[\int d\vec{x}_t\sqrt{\mathcal{P}_1\mathcal{P}_2}\right]$.
\section{Proof of the no-cloning theorem}
Our probability distribution $\mathcal{P}(\vec{x}_t)$ is assumed to evolve according to the Liouville 
equation $\partial_t\mathcal{P}+\nabla\cdot(\vec{v}\mathcal{P})=0$, where $\vec{v}=d\vec{x}_t/dt$ 
denotes the velocity associated with the generalized coordinates in the phase space. Then, the 
derivative of $C_\alpha$ yields
\begin{eqnarray}
\frac{dC_{\alpha}}{dt}&=&-\frac{1}{z}\int d\vec{x}_t \left(\alpha\mathcal{P}_1^{\alpha-1}\mathcal{P}_2^{1-\alpha}
\frac{\partial \mathcal{P}_1}{\partial t}+(1-\alpha)\mathcal{P}_1^{\alpha}\mathcal{P}_2^{-\alpha}
\frac{\partial \mathcal{P}_2}{\partial t}\right)\nonumber\\
&=& \frac{1}{z}\int \!d\vec{x}_t\left[ \alpha\left(\frac{\mathcal{P}_2}{\mathcal{P}_1}\right)^{1-\alpha}
\nabla(\vec{v}\mathcal{P}_1)+(1-\alpha)\left(\frac{\mathcal{P}_1}{\mathcal{P}_2}\right)^\alpha 
\nabla(\vec{v}\mathcal{P}_2)\right]\nonumber\\
&=& \frac{1}{z}\left\{\int d\vec{x}_t \nabla\left(\vec{v}\left(\frac{\mathcal{P}_1}{\mathcal{P}_2}\right)^\alpha
\mathcal{P}_2\right)\right.\nonumber\\
&-& \left. \int \vec{v}\left[\alpha\nabla\left(\frac{\mathcal{P}_2}{\mathcal{P}_1}\right)^{1-\alpha}
\mathcal{P}_1+(1-\alpha)\nabla\left(\frac{\mathcal{P}_1}{\mathcal{P}_2}\right)^\alpha\mathcal{P}_2\right]d\vec{x}_t
\right\}\label{eqn:dct}
\end{eqnarray}
where we put $z=\int d\vec{x}_t\mathcal{P}_1^{\alpha}\mathcal{P}_2^{1-\alpha}$. In the second line, we 
substituted $\partial_t\mathcal{P}_j=-\nabla\cdot(\vec{v}\mathcal{P}_j)$, $(j=1,2)$. The first term 
in the last line vanishes due to the Gauss theorem. Further, the calculations show that the second term 
vanishes. Therefore, the time derivative vanishes, i.e. $d{C}_{\alpha}/dt=0$. Following the line of argument 
addressed in \cite{CNCT,Plastino,Kyoto}, we discuss the particular forms of states before and after 
the cloning process. We need to consider a tri-partite system whose distribution function is assumed 
to be factorized into component distributions as mentioned in (i). 
By the definition provided in Eq.(\ref{eqn:defC}), for the two factorized distributions for systems 
$a$ and $b$, the total distance between two probability distributions $\mathcal{P}_1$ and $\mathcal{P}_2$ 
satisfying Eq.(\ref{eqn:defC}) is equivalent to the sum of the distances between the two systems 
$a$ and $b$,
\begin{eqnarray}
C_\alpha(\mathcal{P}_1(a)\mathcal{P}_1(b),\mathcal{P}_2(a)\mathcal{P}_2(b))
=C_\alpha(\mathcal{P}_1(a),\mathcal{P}_2(a))+C_\alpha(\mathcal{P}_1(b),\mathcal{P}_2(b)).
\end{eqnarray}
Therefore, for the initial ensemble $\mathcal{P}_j=\mathcal{P}_j^{(o)}\mathcal{P}^{(a)}\mathcal{P}^{(r)}$ $(j=1,2)$, 
the distance is attributed to the distance between the original states
\begin{equation}
C_\alpha(\mathcal{P}_1,\mathcal{P}_2)=C_\alpha(\mathcal{P}_1^{(o)},\mathcal{P}_2^{(o)}),
\label{eqn:Ca}
\end{equation}
because we have $C_\alpha(\mathcal{P}^{(a)},\mathcal{P}^{(a)})=0$ and 
$C_\alpha(\mathcal{P}^{(r)},\mathcal{P}^{(r)})=0$. The two premises (ii) and (iii) mentioned 
in Introduction provide the contradiction in question. Indeed, the final distance in the form, 
$\mathcal{Q}_j=\mathcal{P}_j^{(o)}\mathcal{Q}^{(a)}\mathcal{P}_j^{(o)}$, becomes 
\begin{eqnarray}
C_\alpha(\mathcal{Q}_1,\mathcal{Q}_2)=2C_\alpha(\mathcal{P}_1^{(o)},\mathcal{P}_2^{(o)})
+C_\alpha(\mathcal{Q}_1^{(a)},\mathcal{Q}_2^{(a)}).
\end{eqnarray}
This leads to a negative value of $C_\alpha$ if we require that 
$C_\alpha(\mathcal{Q}_1,\mathcal{Q}_2)=C_\alpha(\mathcal{P}_1,\mathcal{P}_2)$ should hold. This, 
however, is in conflict with the positivity property of $C_\alpha$.\\
Further, when we evaluate the $C_\alpha(\mathcal{Q}_1,\mathcal{Q}_2)$ based on the premise (iii), 
we have  
\begin{eqnarray}
C_\alpha(\mathcal{Q}_1,\mathcal{Q}_2)&\geq& -\log\left(\int d\vec{x}^{(o,r)}
\left[\int d\vec{x}^{(a)}_t\mathcal{Q}_1\right]^\alpha\left[\int d\vec{x}^{(a)}_t
\mathcal{Q}_1\right]^{1-\alpha}\right)\nonumber\\
&=& -2\log\left(\int d\vec{x}^{(o)}_t(\mathcal{P}_1^{(o)})^\alpha
(\mathcal{P}_2^{(o)})^{1-\alpha}\right)\nonumber\\
&=& 2C_\alpha(\mathcal{P}_1,\mathcal{P}_2)
\end{eqnarray}
where the first inequality follows from the H{\" o}lder's inequality; the second, from (iii); and 
the last line, from Eq.(\ref{eqn:Ca}). The distance between two different states after copying becomes 
larger than or equal to the twice of the initial one. Hence, in either case, we could show the 
contradiction with Eq.(\ref{eqn:dct}) as long as $\mathcal{P}_1\neq\mathcal{P}_2$ holds for all values of 
$\alpha$. In the present setting, the Liouville dynamics can not allow for classically copying an unknown 
original state to another system without destroying the original system.
\section{Consideration of distance form in terms of a Liouville-like nonlinear evolution equation} 
The Liouville equation must be linear and governs the evolution of a probability density that 
describes a statistical ensemble (in the sense of Gibbs) of dynamical systems, all evolving according to 
the same equations of motion. Such a time-dependent ensemble probability density always evolves according 
to a linear equation (e.g. \cite{van}). Starting from the Liouville theorem, which states the conservation 
of the probability distribution function along the orbit in the phase space, the derived Liouville equation 
cannot have powered probabilities such as $\mathcal{P}^q$, where $q$ may be relevant to deviation from 
linearity. However, it has been pointed out that the linearity of the Liouville equation does not preserve 
statistical independence when the linear combinations of probability distributions 
$\mathcal{Q}=\sum_j c_j\mathcal{Q}_j$ are applied to the associated final states with the marginality 
property\cite{CNCT}. In order to gain more insights into the relation between the dynamics of the 
classical statistical ensemble distribution in copying process and the distance we measure, we here consider 
two cases whose dynamics evolve according to {\it a Liouville-like nonlinear evolution equation}. 
The same reasoning for the classical no-cloning consideration should provide constraints for the 
functional form of the distances and the parameter of the dynamics upon copying.\\

There can be two immediate generalizations of the Liouville equation that break its linearity, 
$\partial_t\mathcal{P}^q+\nabla(\vec{v}\mathcal{P})=0$ and 
$\partial_t\mathcal{P}+\nabla(\vec{v}\mathcal{P}^q)=0$. These forms are reminiscent of the 
nonlinear Fokker-Planck equations, where the probability $\mathcal{P}$ is powered for terms in the 
equations \cite{Frank}. Since the former is equivalent to the latter by a transform 
$\mathcal{P}^q \to \mathcal{P}$, we use the latter in the following consideration. It would be worth 
mentioning the differences between the standard Liouville equation and its nonlinear forms. The Liouville-like 
nonlinear equations may govern the behavior of a time-dependent {\it density} describing a set of real, 
interacting systems instead of an evolution of a probability density associated with a statistical ensemble. 
Indeed, the nonlinearity in the evolution equation for the density constitutes an effective description of the 
interaction between the systems and has relevant applications (e.g. \cite{FrankD}). In this sense, the 
nonlinear Liouville-like equations in our consideration are akin, for instance, to the Vlasov-Poisson equation 
(e.g. \cite{ARP} and references therein). In spite of this difference, we believe that the use of 
the nonlinear Liouville-like equation in the followings serves as a supporting tool for our present discussion.\\
 
First, we consider a distance $G_1$ defined by the following form  
\begin{eqnarray}
G_1=\int d\vec{x}_t \mathcal{P}_1^\alpha h(\eta),\label{eqn:G1}
\end{eqnarray}
where $h$ is a function of the ratio of two different distributions $\eta=\mathcal{P}_2/\mathcal{P}_1$. 
When $\alpha=1$, $G_1$ reduces to the Csisz{\' a}r's f-divergence type, which has already been considered 
in \cite{Plastino}. We are interested in how the exponent $\alpha$ governs the form $h$ when 
$\alpha\neq 1$ and $\alpha\neq 0$. Under the evolution, which can be described by a nonlinear 
generalization of the ordinary Liouville equation, 
\begin{eqnarray}
\partial_t\mathcal{P}+\nabla(\vec{v}\mathcal{P}^q)=0,\label{eqn:nonL1}
\end{eqnarray}
where $q$ is not equal to unity, we consider the constraint on the functional form of $h$ that makes 
the distance $G_1$ time independent. We have 
\begin{eqnarray}
\frac{dG_1}{dt}=\int d\vec{x}_t(\frac{\partial \mathcal{P}_1^\alpha}{\partial t} h+
\mathcal{P}_1^\alpha h^\prime \frac{\partial \eta}{\partial t})\label{eqn:dG1}
\end{eqnarray}
where $h^\prime$ means the derivative with respect to $\eta$. Putting the time derivative of 
$\mathcal{P}_j$ as $-\nabla(\vec{v}\mathcal{P}_j^q)$ from Eq.(\ref{eqn:nonL1}) and from the 
Gauss theorem, we obtain the following expression
\begin{eqnarray}
\frac{dG_1}{dt}&=& -\int d\vec{x}_t \mathcal{P}_1^{\alpha-1}\left[(\alpha h-\eta h^\prime)
\nabla(\vec{v}\mathcal{P}_1^q)+h^\prime\nabla (\vec{v}\mathcal{P}_2^q)\right]\nonumber\\
&=& \int \vec{v}\left[ \mathcal{P}_1^q\nabla\left[\mathcal{P}_1^{\alpha-1}(\alpha h-\eta h^\prime)\right]
+\mathcal{P}_2^q\nabla(\mathcal{P}_1^{\alpha-1}h^\prime)\right]d\vec{x}_t.
\end{eqnarray}
For the rhs. of Eq.(\ref{eqn:dG1}), we find
\begin{eqnarray}
& &\int \vec{v}\left[(\alpha-1) \mathcal{P}_1^{\alpha-2}\left\{\alpha \mathcal{P}_1^q h-
h^\prime(\eta \mathcal{P}_1^q-\mathcal{P}_2^q)\right\}\nabla \mathcal{P}_1\right.\nonumber\\
&+& \left.\mathcal{P}_1^{\alpha-1}\left\{(\alpha-1)\mathcal{P}_1^q h^\prime-
(\eta \mathcal{P}_1^q-\mathcal{P}_2^q)h^{''}\right\}\nabla\eta\right]d\vec{x}_t.
\end{eqnarray}
Because the coefficients of $\nabla P_1$ and $\nabla\eta$ have to vanish, we obtain 
\begin{eqnarray}
\left\{
\begin{array}{r}
\alpha \mathcal{P}_1^q h=(\eta\mathcal{P}_1^q-\mathcal{P}_2^q)h^\prime\\
(\alpha-1)\mathcal{P}_1^qh^\prime=(\eta\mathcal{P}_1^q-\mathcal{P}_2^q)h^{''}.\label{eqn:rel}
\end{array}
\right.
\end{eqnarray}
Simultaneous satisfaction of both relations in Eq.(\ref{eqn:rel}) leads to the ordinary differential 
equation $h^\prime=ch^{\frac{\alpha}{\alpha-1}}$ with the integral constant $c$, which is of the 
Bernoulli type and can be solved by the change of variables $z=h^{\frac{1}{1-\alpha}}$. 
Together with the condition $h(1)=0$, we have 
\begin{eqnarray}
h(\eta)=\left(\frac{c}{1-\alpha}\right)^{1-\alpha}(\eta-1)^{1-\alpha}.\label{eqn:h}
\end{eqnarray}
Note that the nonlinear parameter of the dynamics $q$ does not enter into the functional expression.
Conversely, Eq.(\ref{eqn:h}) must satisfy Eq.(\ref{eqn:rel}). Substituting Eq.(\ref{eqn:h}), e.g.,  
into the first equation of Eq.(\ref{eqn:rel}), we have for any $\eta$
\begin{eqnarray}
\frac{\alpha}{1-\alpha}=\frac{\eta-\eta^q}{\eta-1},
\end{eqnarray}
therefore, we can conclude $q=0$ and $\alpha=1/2$. This case, however, is trivial, indicating that the 
combination of Eq.(\ref{eqn:G1}) and Eq.(\ref{eqn:nonL1}) may be applicable only to the case $q=1$.\\

Let us now consider the case $\alpha=1$ in terms of another possible distance measure. We see that $dG/dt=0$ 
can be achieved robustly under the ordinary Liouville equation 
$\partial_t\mathcal{P}+\nabla (\vec{v}\mathcal{P})=0$ by investigating the following combination:
\begin{eqnarray}
G_2=\int d\vec{x}_t\mathcal{P}_1f(\eta),\quad 
\eta=\frac{\mathcal{P}_2^\beta}{\mathcal{P}_1}
\end{eqnarray}
where $\beta$ is kind of a weighting parameter for the second probability distribution. The similar 
calculation for this provides the result
\begin{eqnarray}
dG_2/dt=\int \vec{v}\left[\beta(\beta-1)\mathcal{P}_2^{q+\beta-2}(\nabla \mathcal{P}_2)f^\prime 
+(\beta \mathcal{P}_2^{q+\beta-1}-\eta \mathcal{P}_1^q)f^{''}\nabla\eta\right]d\vec{x}_t.
\end{eqnarray}
The necessary conditions of $dG_2/dt=0$ are $\beta=0$ or $\beta=1$ from the first term. 
However, $\beta=0$ implies that $G_2$ does not contain $\mathcal{P}_2$, which should be excluded 
from the present consideration. From the vanishing second term, we obtain a relation 
$\beta=(\mathcal{P}_2/\mathcal{P}_1)^{1-q}$. Therefore, as long as $\mathcal{P}_1\neq \mathcal{P}_2$ 
holds, $\beta=1$ yields $q=1$. 
\section{Summary and concluding remarks}
We have shown that the universal cloning machines are incompatible with the conservation 
of information distance that is different from the Csisz{\' a}r's f-divergence under the Liouville 
dynamics. The Chernoff $\alpha$-divergence provides an example that makes the cloning under 
that dynamics infeasible. That is $dC_\alpha/dt= 0$ holds in spite of 
$C_\alpha(\mathcal{Q}_1,\mathcal{Q}_2)\neq C_\alpha(\mathcal{P}_1,\mathcal{P}_2)$ before and after 
the copying process. 
One question still needs to be answered: 
to what extent the present argument is valid? We answer it as follows. 
Since a general form $\mathcal{G}=\int d\vec{x}_t \mathcal{P}_1f(\mathcal{P}_2/\mathcal{P}_1)$ 
satisfies $d\mathcal{G}/dt=0$ under the usual Liouville dynamics \cite{Plastino}, we can see that 
the discussion developed here is also true for the more generic distance measure $G=G(\mathcal{G})$ 
because it immedeately provides 
\begin{eqnarray}
\frac{dG}{dt}=\frac{\partial \mathcal{G}}{\partial t}\frac{\partial G}{\partial \mathcal{G}}=0.
\end{eqnarray}
From this perspective, the Chernoff $\alpha$-distance we employed is only a special case (when the function 
$G$ is logarithmic) that makes the classical counterpart of the quantum no-cloning theorem hold.
\acknowledgments
The author TY acknowledges the hospitality and the visiting grant at the Aston University, where the final 
version of the manuscript was accomplished. The authors thank referees for useful comments upon 
revision. The main part of this paper was presented at the Summer School on Mathematical Aspect of Quantum 
Computing held at Kinki University, $27^{th}-29^{th}$ August 2007.


\begin{thebibliography}{0}
\bibitem{Collin} 
\Name{Collins D. \and Popescu S.}
\REVIEW{Phys. Rev. A}{65}{2002}{032321}.
\bibitem{Woot} 
\Name{Wootters W.K. \and Zurek W.H.}
\REVIEW{Nature}{299}{1982}{802}.
\bibitem{Dieks}
\Name{D. Dieks}
\REVIEW{Phys. Lett. A}{92}{1982}{271}.
\bibitem{Pati} 
\Name{Pati A.K. \and Braunstein S.L.}
\REVIEW{Nature}{404}{2000}{164}. 
\bibitem{CNCT} 
\Name{Daffertshofer A., Plastino A.R. \and Plastino A.}
\REVIEW{Phys. Rev. Lett.}{88}{2002}{210601}.
\bibitem{Mac} 
\Name{Mackey M.C.}
\REVIEW{Rev. Mod. Phys.}{61}{1989}{981}. 
\bibitem{Cs} 
\Name{Csisz{\' a}r I.}
\REVIEW{Studia Math. Hungarica}{2}{1967}{299}.
\bibitem{KL} 
\Name{Kullback S. \and Leibler R.A.}
\REVIEW{Ann. Math. Stat}{22}{1951}{79}.
\bibitem{Plastino} 
\Name{Plastino A.R. \and Daffertshofer A.}
\REVIEW{Phys. Rev. Lett.}{93}{2004}{138701}.
\bibitem{comm} In general, we need not to suppose the system that receives the information of the 
original as {\it blank}, though the device need to erase extra information that the target system 
already possessed.
\bibitem{Cher} 
\Name{Chernoff H.} 
\REVIEW{Ann. Math. Stat.}{23}{1952}{493}.
\bibitem{Bha} 
\Name{Bhattacharyya A.}
\REVIEW{Bull. Calcutta Math. Soc.}{35}{1943}{99}. 
\bibitem{Kyoto} 
\Name{Plastino A.R.}
\REVIEW{Prog. Theor. Phys. Suppl.}{162}{2006}{173}.
\bibitem{van}
\Name{van Kampen N.G.}
\Book{Stochastic Processes in Phyics and Chemistry 3rd ed.}
\Publ{North Holland}
\Year{2007}.
\bibitem{Frank}
\Name{Frank T.D.}
\Book{Nonlinear Fokker-Planck Equation}
\Publ{Springer}
\Year{2005}.
\bibitem{FrankD} 
\Name{Frank T.D, Daffertshofer A., Peper C.E., Beek P.J., \and Haken H.}
\REVIEW{Physica D}{144}{2000}{62}.
\bibitem{ARP} 
\Name{Plastino A.R, Giordano C., Plastino A., \and Casas M.}
\REVIEW{Physica A}{336}{2004}{376}.
\end{thebibliography}
\end{document}